\documentclass[11pt]{article}
\usepackage{hyperref}
\pdfoutput=1

\begin{document}
\title{Influences of initial streamwise rotation of a droplet under a uniform flow}
\author{Eric K. W. Poon\\
        \vspace{6pt}
        Institute of High Performance Computing, Singapore $138632$\\
        and\\
        Department of Mechanical Engineering, University of Melbourne,\\
        Parkville, Victoria $3010$, Australia\\
        \\\vspace{6pt}
        Shaoping Quan and Jing Lou\\
        \vspace{6pt}
        Institute of High Performance Computing, Singapore $138632$\\
        \\\vspace{6pt}
        Andrew S. H. Ooi\\
        \vspace{6pt}
        Department of Mechanical Engineering, University of Melbourne,\\
        Parkville, Victoria $3010$, Australia}

\maketitle

\begin{abstract}
A spherical droplet is given an initial rotation in the streamwise direction and is impulsively accelerated by a uniform free stream. Numerical results for the deformation and dynamics of the droplet are obtained by utilising a finite volume staggered mesh method with a moving mesh interface tracking scheme. The fluid dynamics videos of the droplet are presented in the Gallery of Fluid Motion, $2010$. By initially rotating the droplet in the streamwise direction, the droplet deforms differently depending on the non-dimensional rotation rate, $\Omega^*$. The families of droplet shape are, in ascending $\Omega^*$, biconvex, convex-concave and biconcave. While the biconvex and convex-concave families are due to the compression by the combined vortex ring $($across the lee side of the interface$)$. The biconcave family is formed because of the huge surface tension at the droplet edge that restrains further deformation there. In a special case, the biconcave family releases the combined vortex ring from the droplet. The release of the vortex ring is observed to be at $Re$ one order of magnitude lower than a similar mechanism observed for the flow past a solid sphere. This release of the vortex ring has led to a series of events and a reduction in $C_D$.
\end{abstract}

\section{Introduction}
The two videos for the initially streamwise rotating droplet under a uniform cross flow are
\href{http://arxiv.org/src/1010.3081v1/anc/video1.mpg}{Video1}
and
\href{http://arxiv.org/src/1010.3081v1/anc/video2.mpg}{Video2}.

In the link videos, the computation is performed using a finite volume staggered mesh method. The interface is located using a moving mesh interface tracking $($MMIT$)$ scheme \cite{Quan2007}. In the MMIT scheme, the interface is represented by a triangular surface element and is of zero-thickness. The benefits of ulitising the MMIT method are the total mass of each phase is naturally conserved and it allows direct implementation of the jumps in fluid properties and boundary conditions across the interface.

The droplet is travelling at a single initial Reynolds number $(Re = 40)$ in this work. Initial Weber numbers, $We = 4$ and $40$, are considered. The viscosity ratio $(\lambda = 50)$ and density ratio $(\eta = 50)$ are set constant throughout the simulation. The effects of streamwise rotation on the deformation and dynamics of the droplet is investigated at non-dimensional rotation rates $0 < \Omega^* \leq 1$. Although simulations have been performed at $\Omega^*$ from $0$ to $1$, the linked movies show only the cases with $Re = 40$, $We = 40$, $\lambda = \eta = 50$, $\Omega^* = 0.2, 0.6, 1$ and $Re = 40$, $We = 4$, $\lambda = \eta = 50$, $\Omega^* = 1$ to highlight the important founding.

The biconvex family shape is observed at $Re = 40$, $We = 40$, $\lambda = \eta = 50$, $\Omega^* = 0.2$. At initial condition, a small vortex ring is observed in the near-wake, while a larger vortex ring is shown inside the droplet. Once the droplet is released downstream, the two vortex rings combine to form a single large vortex ring across the lee side of the interface. The motion of the vortex ring compresses the lee side of the droplet, whereas the upwind side of the droplet is pushed downstream by the free stream inertia. As a result, the droplet turns into an elliptical shape and then becomes biconvex. It is also evident in the movies that the drag coefficient, $C_D$, is steadily increasing throughout the simulation. The possible physical explanation of the increase in $C_D$ is attributed an increase in frontal area of the droplet, which makes the droplet more vulnerable to the free stream inertia. Therefore increasing the droplet centroid velocity and thus the increases $C_D$.

A larger vortex ring is observed in the near-wake for $Re = 40$, $We = 40$, $\lambda = \eta = 50$, $\Omega^* = 0.6$. The size of the combined vortex ring is therefore increased after the droplet is released downstream. The larger combined vortex ring not only spins the droplet further away from the rotation axis. It also further compresses the lee side of the droplet such that a concave depression is formed at the late stages. Due to the droplet spinning further away from the streamwise axis, the frontal area is larger as compared to $\Omega^* = 0.2$ and thus a higher $C_D$ is observed.

A further increase in $\Omega^*$ results in an even larger combined vortex ring across the interface. As a result, the droplet becomes very thin and the large curvature at the edge restrains further deformation there. On the other hand, the centre part of the droplet flattens and surface tension there is relatively small. Thus, the centre part of the droplet is kept compressing by the vortex ring and the free stream. The droplet turns into a biconcave shape as a result.

In the movies, the deformation and dynamics of the droplet at $Re = 40$, $We = 4$, $\lambda = \eta = 50$, $\Omega^* = 1$ are also presented. The droplet deforms into a biconcave shape at the early stages. However, as the surface tension at the edge increases, the droplet expansion slows down. The increase in droplet centroid velocity is therefore slowed down and decreases the time derivative of the droplet centroid velocity. As a result, the drag coefficient begins to decrease. The other impact of increase surface tension at the edge is a concave depression on the upwind side of droplet. The concave depression leads to the release of the combined vortex ring downstream, which the release of vortex ring only happens at a $Re$ of at least one order of magnitude higher for the flow past a sphere. The release of the vortex ring reduces the circulation at the droplet which results in a drag reduction \cite{Howe2001}. The other causes for the drag reduction is attributed to the formation of a vortex ring on the upwind side of the droplet. As the droplet shrinks, this vortex enlarges and pushes the droplet upstream. The droplet centroid velocity is slowed down dramatically which further reduces $C_D$.

The authors would like to acknowledge the support by Agency for Science, Technology and Research $($A*Star$)$, Singapore for this work.

\bibliographystyle{plain}
\bibliography{galleryOfFluidMotion}

\end{document}